\documentclass[aps,prl,twocolumn,groupedaddress]{revtex4-1}

\usepackage{amsmath}

\usepackage{graphicx}

\DeclareGraphicsExtensions{.pdf,.eps,.png,.jpg,.mps}

\usepackage{epstopdf}

\usepackage{threeparttable}

\usepackage{color}

\begin{document}

\title{Broadband optical parametric amplification by two-dimensional semiconductors}

\author{Chiara Trovatello$^{1,2}$, Andrea Marini$^{3}$, Xinyi Xu$^{2}$, Changhwan Lee$^{2}$, Fang Liu$^{4}$, Nicola Curreli$^{2,5}$, Cristian Manzoni$^{6}$, Stefano Dal Conte$^{1}$, Kaiyuan Yao$^{2}$, Alessandro Ciattoni$^{7}$, James Hone$^{2}$, Xiaoyang Zhu$^{4}$, P. James Schuck$^{2}$, and Giulio Cerullo$^{1,6}$}

\affiliation{$^1$Dipartimento di Fisica, Politecnico di Milano, Piazza L. da Vinci 32, I-20133 Milano, Italy}

\affiliation{$^2$Department of Mechanical Engineering, Columbia University, 10027, New York, NY, USA}

\affiliation{$^3$Department of Physical and Chemical Sciences, University of L'Aquila, Via Vetoio, 67100 L'Aquila, Italy}

\affiliation{$^4$Department of Chemistry, Columbia University, 10027, New York, NY, USA}

\affiliation{$^5$Graphene Labs, Istituto Italiano di Tecnologia, via Morego 30, 16163 Genova, Italy}

\affiliation{$^6$IFN-CNR, Piazza L. da Vinci 32, I-20133 Milano, Italy}

\affiliation{$^7$CNR-SPIN, c$/$o Dip.to di Scienze Fisiche e Chimiche - Via Vetoio - 67010 Coppito (AQ), Italy}

\begin{abstract}
\textbf{Optical parametric amplification is a second-order nonlinear process whereby an optical signal is amplified by a pump via the generation of an idler field. It is the key ingredient of tunable sources of radiation that play an important role in several photonic applications. This mechanism is inherently related to spontaneous parametric down-conversion that currently constitutes the building block for entangled photon pair generation, which has been exploited in modern quantum technologies ranging from computing to communications and cryptography. Here we demonstrate single-pass optical parametric amplification at the ultimate thickness limit; using semiconducting transition-metal dichalcogenides, we show that amplification can be attained over a propagation through a single atomic layer. Such a second-order nonlinear interaction at the 2D limit bypasses phase-matching requirements and achieves ultrabroad amplification bandwidths. The amplification process is independent on the in-plane polarization of the impinging signal and pump fields. First-principle calculations confirm the observed polarization invariance and linear relationship between idler and pump powers. Our results pave the way for the development of atom-sized tunable sources of radiation with applications in nanophotonics and quantum information technology.}
\end{abstract}

\maketitle

\section{Introduction}

\begin{figure*}[t!]
\centerline{\includegraphics[width=190mm]{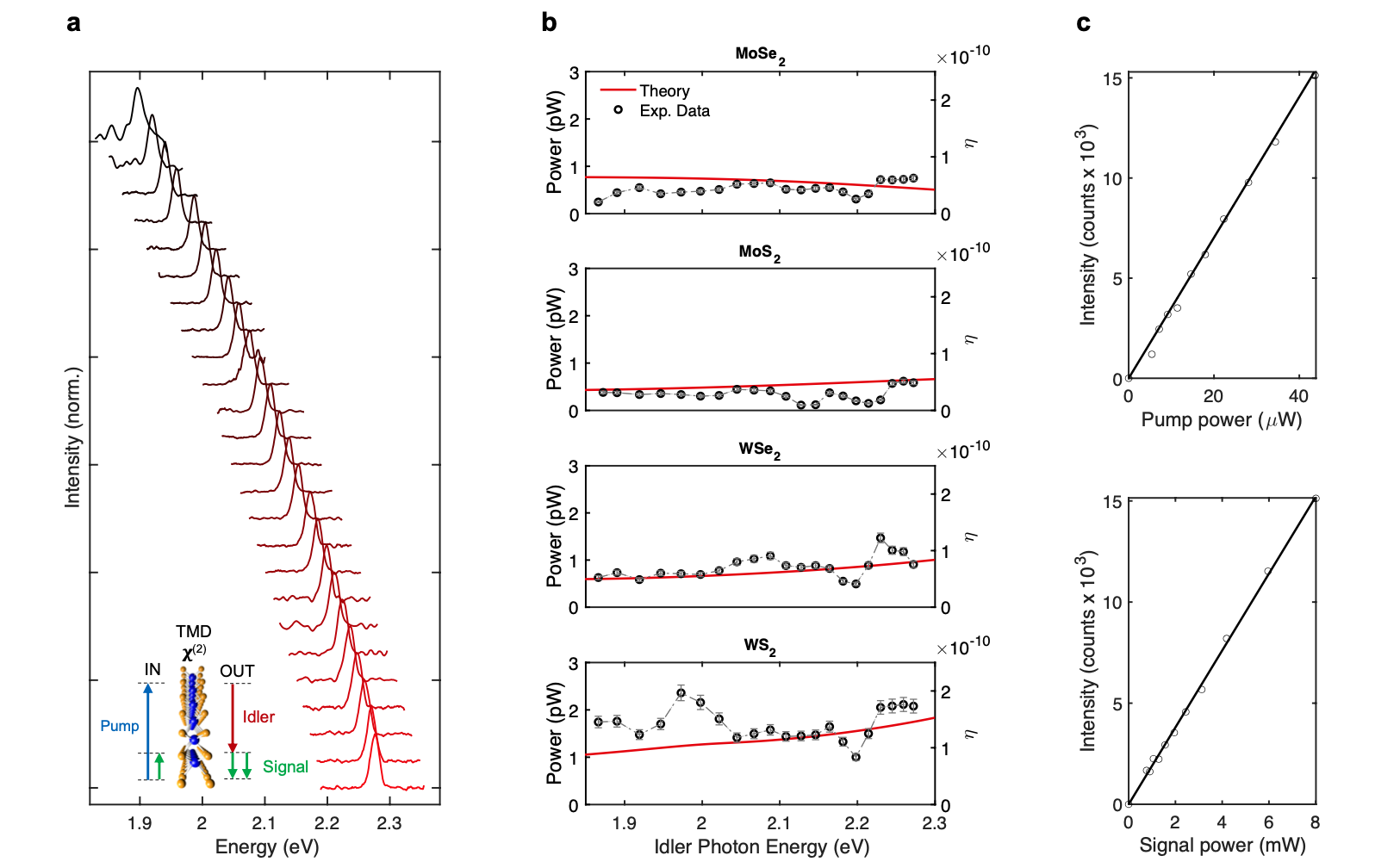}}
\caption{{\bf Broadband OPA in 1L-TMDs. (a)} Normalized tunable idler spectra measured on 1L-MoSe$_2$; {\bf (b)} absolute idler power/efficiency $\eta$ as a function of the idler photon energy $\hbar\omega_{\rm i}$ measured on the four semiconducting TMDs (black dots), and calculated corresponding theoretical efficiencies divided by a factor of 2.5 (continuous red lines). The power of the pump is fixed at $48.3$ $\mathrm{\mu}$W while the signal power is fixed to $12.7$ mW. The pump polarization is linear and vertical, while the signal polarization is linear and horizontal. Pump and signal beams are focused down to spot sizes of $\sim 1$ $\mathrm{\mu}$m and $\sim 2$ $\mathrm{\mu}$m, respectively; {\bf (c)} idler intensity linear dependence on pump and signal powers measured on 1L-MoSe$_2$.}
\label{Fig:efficiency}
\end{figure*}

When an intense electromagnetic wave interacts with matter, the induced polarization contains terms that are nonlinear in the driving field, giving rise to a plethora of physical phenomena and photonic applications such as frequency conversion \cite{Franken1961}, all-optical signal processing \cite{Stegeman1996}, and non-classical sources of radiation \cite{Kwiat1995}. In particular, parametric down-conversion (PDC) is the annihilation of a high-energy pump photon $\hbar\omega_{\rm p}$ into a pair of lower energy photons (signal with $\hbar\omega_{\rm s}$ and idler with $\hbar\omega_{\rm i}$) according to energy conservation $\hbar\omega_{\rm p} = \hbar\omega_{\rm s} + \hbar\omega_{\rm i}$. PDC is a second-order nonlinear process which underpins optical parametric amplification (OPA) and oscillation (OPO), which are exploited in tunable sources of coherent radiation \cite{Giordmaine1965,Yariv1966} and for the generation of entangled photons and squeezed states of light \cite{Wu1987}. Current OPA/OPO setups exploit anisotropic crystals with broken inversion symmetry that enable momentum conservation in the PDC process by the birefringence phase-matching (PM) technique \cite{Giordmaine1962} or alternatively by quasi-PM, which exploits the extra-momentum provided by the periodic reversal of the crystal orientation to attain momentum conservation \cite{Vodopyanov2004,Canalias2007}. In conventional nonlinear materials, the PM requirement limits the signal bandwidth over which the OPA/OPO process can occur.

Transition metal dichalcogenides (TMDs) are layered semiconducting materials which can be reduced to single-layer (1L) thickness due to the weak van der Waals interlayer forces. 1L-TMDs are 2D semiconductors which, despite the vanishing ($<1$nm) thickness, exhibit very strong light-matter interaction due to quantum confinement effects\cite{Chernikov2014,Qiu2013}. This class of materials also possess second order susceptibility $\chi^{(2)}$ which is orders of magnitude higher than in conventional nonlinear materials\cite{Yin2014,Kumar2013,Li2013,Malard2013,Janisch2014,Majumdar2015,Le2016,Fryett2017,Yu2017,Ciattoni2018,Autere2018,Woodward2016,Sanyatjoki2017,Mennel2018,Autere20182,Hsu2014,Li2016,Jiang2019,Janish2014,Mennel2019}. \\ 
Here we show that 1L-TMDs enable single-pass collinear OPA at the atomic scale without the PM constraint, thanks to the 2D nonlinear interaction in a medium with vanishing thickness that does not introduce any dispersion-induced phase mismatch. We demonstrate this process in the most studied group VI 1L-TMDs, namely MoSe$_2$, MoS$_2$, WSe$_2$, and WS$_2$. We further demonstrate that OPA is insensitive to the crystal orientation and to the polarization directions of pump and signal waves and that the power of the measured idler wave scales linearly with the pump and the signal power. Our experimental findings are fully supported by first-principle theoretical calculations based on the tight-binding model \cite{LWY2013} and on the Bloch independent electron dynamics \cite{Ciattoni2018}, indicating that the observed phenomena are a direct consequence of the D$_{\rm 3h}$ group symmetry of such crystals. We further demonstrate that artificial AA stacking of 1L-TMDs allows quadratic efficiency scaling with the number of layers while still remaining in a deeply sub-wavelength thickness regime.

\section{Results and Discussion}

The MoSe$_2$, MoS$_2$, WSe$_2$, and WS$_2$ monolayers are prepared using a gold-assisted exfoliation technique (see Methods), which enables the fabrication of large area (mm size) samples. They are subsequently transferred on top of $500$ $\mathrm{\mu}$m thick SiO$_2$ substrates. Samples are illuminated by a 40X reflective objective with two collinear and synchronized femtosecond laser beams, supplied by a Ti:Sapphire oscillator and an OPO operating at an $80$ MHz repetition rate. Pump and signal beams are overlapped in space and synchronized in time at the sample plane. The nonlinear emission is collected in a back scattering geometry onto a Silicon-EMCCD camera (see Methods).

As sketched at the bottom left Fig.\ref{Fig:efficiency}a, in an OPA process part of the pump photons are annihilated into pairs of signal and idler photons, following the energy conservation constraint. We chose to measure the idler beam, which is the difference frequency (DF) between pump and signal, as a fingerprint of the OPA process due its background free detection. \\

\textbf{Efficiency}

\begin{figure*}[t!]
\centerline{\includegraphics[width=170mm]{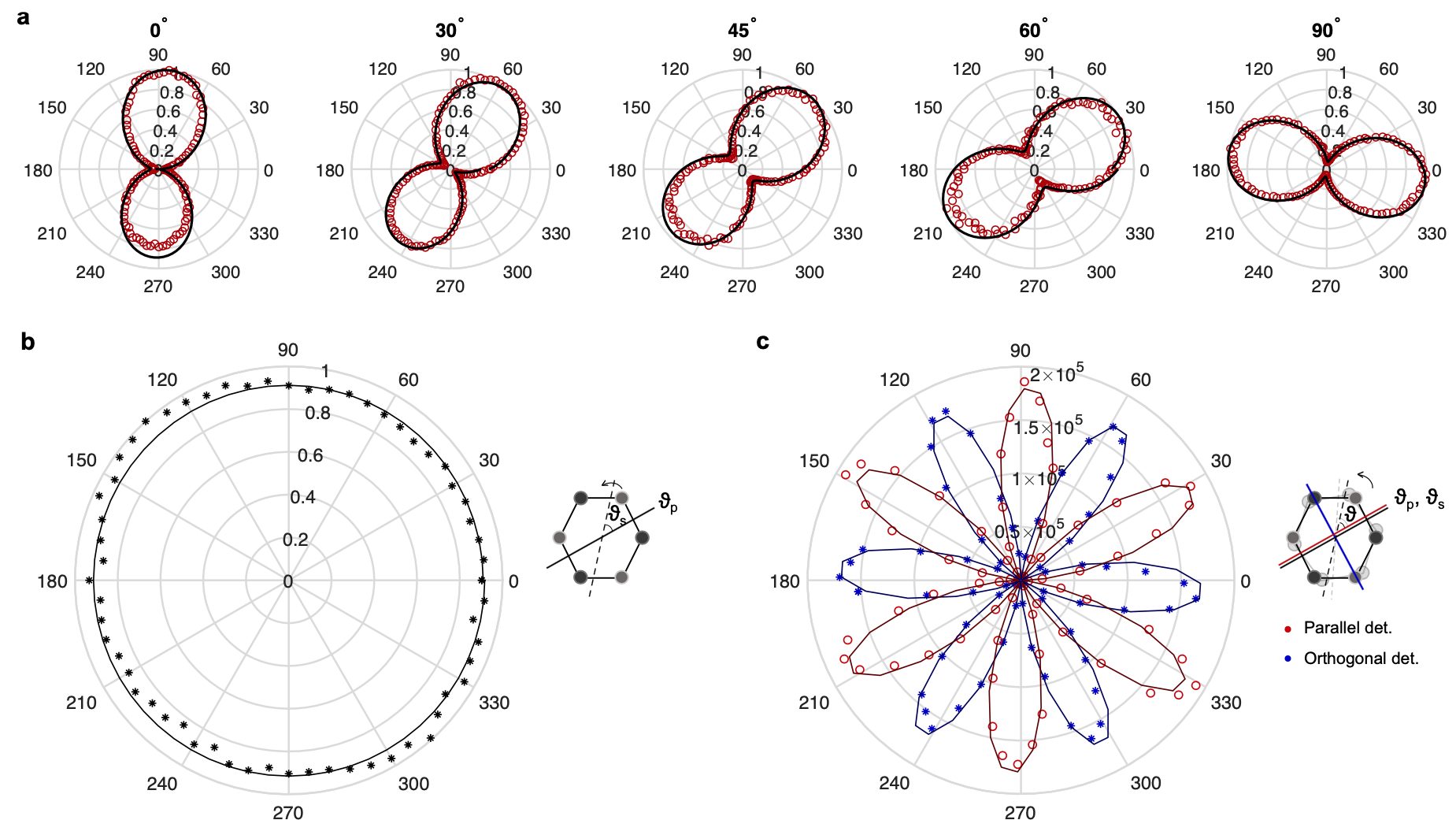}}
\caption{{\bf Idler polarization.} {\bf (a)} Polar plots of the measured polarization emission of the emitted idler beam at $2.15$ eV (dots) and fitting functions (lines) $f \propto cos^2(\theta+\phi)$, $\phi$ is the offset angle. The signal polarization is linear and horizontal ($0^\circ$) while the pump polarization is tuned at the discrete values $0^\circ$, $30^\circ$, $45^\circ$, $60^\circ$, $90^\circ$. {\bf (b)} Polar plot of the idler spectrum intensity (dots) centered at $2.15$ eV as a function of $\theta_{\rm s}-\theta_{\rm p}$, and fitting constant function (continuous line); {\bf (c)} Polar plot of the idler intensity as a function of the crystal's azimuthal angle $\theta$ (dots). Red and blue colors indicate the parallel and perpendicular components of the idler intensity with respect to pump and signal beams, which are set to be linearly polarized to $0^\circ$. The fitting function (continuous line) for both detection configurations is $f \propto cos^2(3\theta+\phi)$, $\phi$ is the offset angle, relative to the armchair direction.}
\label{Fig:emission}
\end{figure*}

Figure \ref{Fig:efficiency} demonstrates the broadband tunability and quantifies the efficiency of the OPA process in 1L-TMDs. Figure \ref{Fig:efficiency}a shows the normalized idler spectra measured on a monolayer of MoSe$_2$ in a broad photon energy range from 1.9 eV to 2.3 eV. The finite spectral window is solely limited by the signal beam tunability. Figure \ref{Fig:efficiency}b shows four panels reporting the absolute values of the effective generated idler power (black dots) measured across the four semiconducting 1L-TMDs, already accounting for all the estimated experimental losses (see Methods). The $500$ $\mathrm{\mu}$m SiO$_2$ substrate does not provide any appreciable nonlinear signal in the observed spectral window under the same experimental conditions (see Supplementary Information). Although excitons have been shown to play a relevant role in the linear and nonlinear optical properties of 1L-TMDs \cite{Ugeda2014,Selig2016}, the almost flat spectral efficiency of idler signal observed for all the 1L-TMDs (see Fig.\ref{Fig:efficiency}b) implies that they are not playing a significant role in the OPA process for the considered photon energies and intensities of pump and signal fields. This result contrasts the resonant behavior of the second harmonic generation process observed around the excitons\cite{Wang2015,Seyler2015}. Indeed, our theoretical calculations (red solid lines in Fig. \ref{Fig:efficiency}b, see Methods and Supplementary Information) reproduce with remarkable accuracy such a flat,  broadband and almost featureless efficiency. This is attributed to independent interband dynamics in 1L-TMDs. The observed ultrabroad bandwidth of the OPA process is related to the vanishing thickness of the nonlinear material, which eliminates any phase mismatch\cite{Ciattoni2018}. Owing to the D$_{\rm 3h}$ group symmetry of 1L-TMDs, the second order nonlinear susceptibility tensor is characterized by only one parameter $\chi^{(2)}(\omega_{\rm p},\omega_{\rm s})$ \cite{Li2013,ShenBook}. Our theoretical calculations indicate that the broadband spectrum of the OPA process ensues from a nonresonant effective $|\chi^{(2)}(\omega_{\rm p},\omega_{\rm s})|\simeq 2 - 8 \times 10^{-10}$ m$/$V for the 1L-TMDs (see Methods and Supplementary Information), which is of the same order of magnitude of the resonant $|\chi^{(2)}|$ of typical bulk semiconductors like GaAs \cite{Bergfeld2003}. Figure \ref{Fig:efficiency}c shows a linear dependence of the idler intensity with pump and signal power measured on 1L-MoSe$_2$ (see full lines in Fig. \ref{Fig:efficiency}c), in agreement with the theory that reproduces accurately such dependence in the undepleted pump approximation and weak excitation regime (see Methods and Supplementary Information). The other TMDs - WS$_2$, WSe$_2$ and MoS$_2$ - show a similar behavior (see Supplementary Information).\\

\textbf{Polarization}

\begin{figure*}[t!]
\centerline{\includegraphics[width=140mm]{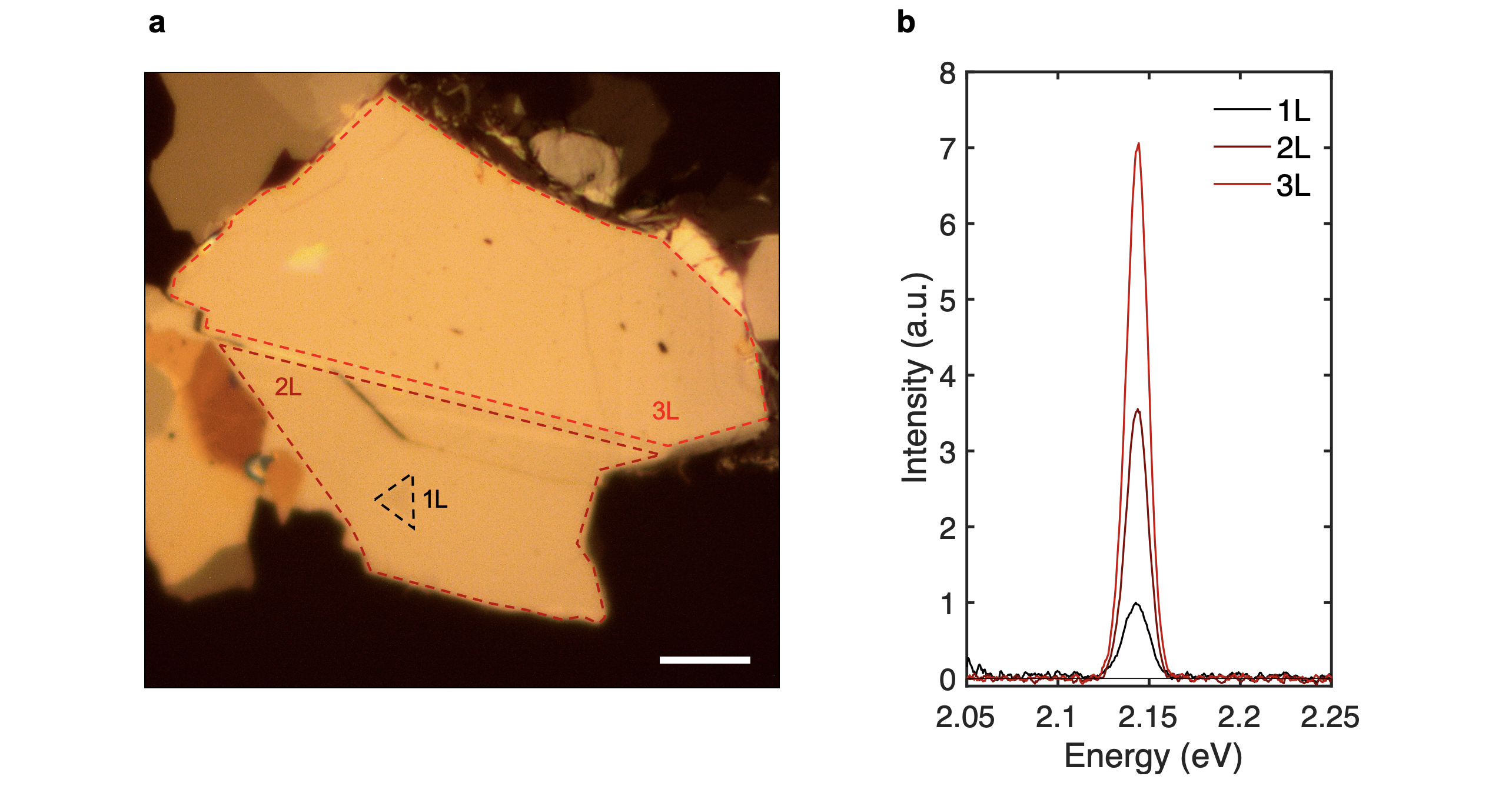}}
\caption{{\bf Efficiency scaling with AA stacking.} {\bf (a)} Micrograph of the WS$_2$ AA stack sample, the scale bar is $10\mu m$; (b) Idler intensity as a function of the number of layers. The power of the pump is fixed at $45$ $\mathrm{\mu}$W while the signal power is fixed at $10$ mW. Pump and signal beams have linear and horizontal polarization, and they are focused down to a spotsize of $\sim 1$ $\mathrm{\mu}$m and $\sim 2$ $\mathrm{\mu}$m, respectively. }
\label{Fig:AAstack}
\end{figure*}

The theoretical analysis of the OPA process in such 1L-TMDs, in addition to providing the second order nonlinear susceptibility tensor, reveals the polarization dependence of the amplified signal field and generated idler field. The D$_{\rm 3h}$ group symmetry implies that the OPA efficiency does not depend on the absolute and the mutual polarizations of signal and pump fields (see Supplementary Information). Conversely, the polarization of the idler signal is fixed by the pump and signal in-plane polarization angles $\theta_{\rm p}$ and $\theta_{\rm s}$, respectively, with respect to the armchair crystallographic axis and follows the rule $\theta_{\rm i}=\pi/2-\theta_{\rm p}-\theta_{\rm s}$, which we confirm experimentally. Figure \ref{Fig:emission}a reports polar plots of the idler generation efficiency for different pump polarization angles for 1L-WSe$_2$. The signal polarization is kept horizontal ($\theta_{\rm s} = 0^\circ$) while the pump polarization is rotated by a half-waveplate at five discrete angles: $\theta_{\rm p} = 0^\circ$, $30^\circ$, $45^\circ$, $60^\circ$, $90^\circ$. The polar measurements depicted in Fig.\ref{Fig:emission}a are taken rotating an analyzer  in front of the the detector. Red circles represent the measured experimental data while black lines are the obtained fitting curves using the function cos$^2(\theta+\phi)$, where $\theta$ and $\phi$ stand for the polarizer angle and an offset angle (reflecting the alignment of the armchair direction), respectively, confirming the theoretical prediction of linearly polarized idler signal and the dependence of $\theta_{\rm i}$ over $\theta_{\rm p}$. 

Figure \ref{Fig:emission}b shows the normalized total emitted idler intensity at $2.15$ eV as a function of the pump-signal polarization angle ($\theta_{\rm s}-\theta_{\rm p}$) measured on 1L-WSe$_2$. The pump polarization is linear and horizontal (0$^\circ$) while the signal polarization is rotated from 0$^\circ$ to 360$^\circ$. The data display an efficiency independent of the relative angle between pump and signal polarization directions, in agreement with theoretical predictions. The dependence of the idler intensity on the crystallographic orientation is reported in Fig.\ref{Fig:emission}c for parallel and perpendicular directions of the polarizer axis in detection. These polar plots reproduce results previously reported with second-harmonic generation (SHG) confirming the six-fold symmetric second-order nonlinear tensor of the material \cite{Li2013}. Since they belong to the same $D_{3h}$ group symmetry, the other TMDs - WS$_2$, MoS$_2$ and MoSe$_2$ - display qualitative behavior identical to those in Fig.\ref{Fig:emission}b-c under the same experimental conditions, as shown in the Supplementary Information for 0$^\circ$-0$^\circ$ and 0$^\circ$-90$^\circ$ signal-pump input polarization configurations and for SHG polar measurements.\\

\textbf{N$^2$ gain enhancement in artificial AA stack}

In our operating regime, the gain of the second-order parametric process scales quadratically with the thickness of the nonlinear medium. Therefore, the effective OPA gain, in principle, could be boosted by increasing the number of TMD layers in the sample. Unfortunately, in the natural vertical stacking of the most commonly used semiconducting multilayer TMDs (2H polytype), each constituting layer is rotated by 180$^\circ$ with respect to its next neighbours, forming the so called AB stack\cite{He2014}. As a direct consequence, the inversion symmetry in TMD samples with even number of layers is restored ($|\chi^{(2)}|=0$), preventing the observation of any second-order nonlinear process. One possibility to increase the OPA effective gain, preserving the broken inversion symmetry ($|\chi^{(2)}|\neq0$), is to vertically stack several monolayers with interlayer $\sim0^\circ$ crystal angle alignment, forming the so called AA stack\cite{He2014}. Because of the atomic thickness of TMDs, many layers can be stacked on top of each other while still maintaining a deeply sub-wavelength thickness and avoiding PM constraints. In an ideal lossless case, the OPA efficiency is expected to scale as the square of the number of layers N\cite{Zhao2016}. We have performed a proof-of-principle demonstration of this concept for 1-3 layers of manually AA stacked WS$_2$, shown in Fig.\ref{Fig:AAstack}a. We can distinguish 1L, 2L and 3L-WS$_2$ regions, sitting on top of a $\sim30nm$ hBN flake (see Methods). The whole AA stack is transfered on top of a $500\mu m$ thick SiO$_2$ substrate. Figure \ref{Fig:AAstack}b shows the emitted idler spectrum at 2.15eV, as a function of the number of WS$_2$ layers. Pump and signal photon energies are set to 3.1eV and 0.95eV, respectively, and both beams have parallel and horizontal polarizations with respect to the sample plane. The underlying hBN flake does not give any appreciable second order nonlinear signal within our experimental conditions, since $|\chi^{(2)}|$ for bulk hBN is 2-3 orders of magnitude smaller than for 1L-TMDs\cite{Li2013}. The measured idler intensity indeed scales nearly quadratically with the layer number. The observed deviation from the $N^2$ trend is due to the absorption of the 3.1eV pump beam which lies above the bandgap of WS$_2$, $\alpha_p\sim12\%$, and to the partial reabsorption of the emitted idler field, $\alpha_i\sim3\%$ (see Supplementary Information). Such absorption effect can be easily avoided by tuning the pump photon energy below the bandgap of the TMD. AA stacking thus offers a clear route for scaling the efficiency of the OPA process in TMDs in view of their applications in photonic devices. Similarly, non-centrosymmetric 3R-oriented TMD crystals can be used, which are directly grown via CVD\cite{Zhao2016}.

\section{Conclusions}

Our experimental and theoretical investigations provide the first evidence of OPA by a single pass through 1L-TMDs. We find that the amplification efficiencies of MoS$_2$, MoSe$_2$, WS$_2$, and WSe$_2$ are comparable and that their spectral dependance is flat, thus enabling broadband functionality. The measured broadband behavior and efficiency of the nonlinear process are fully reproduced by first-principle calculations based on the Bloch independent electron dynamics, thus implying that excitons do not play a significant role in the OPA process for the considered photon energies and intensities. Furthermore, the OPA process in 1L-TMDs is observed across an ultrabroad spectral range, bypassing PM costraints: this is generally unattainable in standard photonic materials owing to destructive interference produced by chromatic dispersion. Artificial stacking of AA aligned TMD monolayers provides a route for quadratic scaling of the efficiency with layer number, while still maintaining the ultrabroad bandwidth enabled by the deeply sub-wavelength regime. Our results shed light on second-order parametric processes in 1L-TMDs, paving the way to the scaling and the integration of 2D materials in future photonic applications, such as 2D all optical amplifiers, single-photon nanoemitters and integrated sources of entangled photons.

\vspace{1cm}

{\bf Acknowledgments}

C.T., G.C. and N.C. acknowledge the project SONAR, which has received funding from the European Union’s Horizon 2020 research and innovation programme under the Marie Sk\l odowska-Curie grant agreement no. 734690. G.C. acknowledges support by the European Union Horizon 2020 Programme under Grant Agreement No. 785219 Graphene Core 2. A.M. acknowledges support from the Rita Levi Montalcini Fellowship (grant number PGR15PCCQ5) funded by the Italian Ministry of Education, Universities and Research (MIUR). A.C. aknowledges PRIN 2017 PELM (grant number 20177PSCKT). XYZ acknowledges partial support by the National Science Foundation (NSF) grant DMR-1809680 for the development of the exfoliation technique. 
This research was supported in part by the Department of Energy (DOE) Office of Energy Efficiency and Renewable Energy (EERE) Postdoctoral Research Award under the EERE Solar Energy Technologies Office administered by the Oak Ridge Institute for Science and Education (ORISE) for the DOE. ORISE is managed by Oak Ridge Associated Universities (ORAU) under DOE contract number DE-SC00014664. All opinions expressed in this paper are the author's and do not necessarily reflect the policies and views of DOE, ORAU, or ORISE. P.J.S., J.H., and K.Y. acknowledge support from Columbia University's Fu Foundation School of Engineering and Applied Science Interdisciplinary Research Seed (SIRS) Funding program. Portions of this work are supported as part of Programmable Quantum Materials, an Energy Frontier Research Center funded by the U.S. Department of Energy (DOE), Office of Science, Basic Energy Sciences (BES), under award DE-SC0019443. 

\vspace{1cm}

{\bf Competing financial interests}

The authors declare no competing financial interest.

\newpage

\section{Methods}

{\bf Sample preparation}.

The monolayers MoS$_2$, WS$_2$, MoSe$_2$, WSe$_2$ on SiO$_2$ substrate are prepared from bulk WS$_2$, MoSe$_2$, WSe$_2$ (HQ graphene) and MoS$_2$ (SPI Supplies) single crystals, using a gold assisted exfoliation technique published previously\cite{Liu2019}. Briefly, a layer of gold is prepared on top of the bulk crystals. The gold layer is exfoliated away with a thermal release tape (semiconductor corp.), carrying a large piece of TMD monolayer on the contact surface, and is transferred onto a SiO$_2$ substrate. The thermal release tape is removed by heating up at 130$^\circ$C.  The tape residues are cleaned by acetone and O$_2$ plasma treatment. The gold layer is dissolved with a gold etchant solution, which is made from mixing KI(99.9\%, Alfa Aesar) and I$_2$ (99.99\%, Alfa Aesar) in DI water. The exfoliated monolayers have a clean surface and strong PL comparable with the monolayers from conventional scotch tape exfoliation\cite{Liu2019}.

The AA stack WS$_2$ sample is prepared employing a dry, contamination-free transfer procedure in which flakes are assembled into a perfectly oriented heterostructure ($\sim0^\circ$ crystal angle mismatch). A $\sim$30nm thick hBN crystal is used as a sacrificial layer to allow the pick-up of 1L-WS$_2$ crystals. hBN and 1L-WS$_2$ crystals are separately exfoliated on SiO$_2$ substrates through mechanical and gold-assisted large area exfoliation respectively\cite{Liu2019}.
A polypropylene carbonate (PPC) polymer on a polydimethylsiloxane (PDMS) elastomer stamp is used to peel a hBN flake directly off the substrate under an optical microscope. To do this, the stamp is attached to a three-axis (XYZ) manipulator with the flakes facing towards the sample. With the same procedure, the hBN flake is positioned over the 1L-WS$_2$ flake; as the stamp is transparent, one can see the sample through it, thus it is possible to align the edges of hBN and WS$_2$. The two flakes are then brought into contact while the whole system is heated to promote adhesion between the flakes. The two flakes develop a strong van der Waals adhesion, and, as a result, the 1L-WS$_2$ is torn and peeled off its substrate as the elastomer stamp is raised. A second 1L-WS$_2$ flake is vertically added to the stack following the same procedure. In order to guarantee $\sim0^\circ$ crystal angle mismatch among all the stacked layers, the hBN/1L-WS$_2$ stack is exclusively translated on top of the remaining large area 1L-WS$_2$. The third 1L-WS$_2$ flake is finally peeled off its substrate from hBN/1L-WS$_2$/1L-WS$_2$. Once the stacking procedure is completed, the pristine $500\mu m$ thick SiO$_2$ wafer acceptor substrate is positioned onto the sample's XYZ manipulator and the stack is deposited on it for further characterization.
The experimental setup employed to transfer the 2D crystals comprises a  custom built Signatone CM300 transfer station.\\

{\bf Experimental Setup}.

The laser source is a Ti:Sapphire oscillator (Coherent Chameleon Ultra II), which emits 150fs pulses at 1.55eV, with a repetition rate of 80MHz and an average output power of 4W. The oscillator pumps an OPO, emitting a tunable signal from 0.83eV to 1.23eV. The pump beam at 3.1eV is obtained by frequency doubling the laser output by Type I 2-mm-thick beta barium borate (BBO) crystal. The 1.55eV laser is focused on the BBO with a lens of focal distance f=50mm, and the generated second harmonic (SH) at 3.1eV is collimated with a f=75mm lens. Since the 1.55eV beam has a linear horizontal polarization, the generated 3.1eV beam has a linear vertical polarization. The fundamental beam is filtered out by a short pass 500nm (2.48eV) filter, placed after the nonlinear crystal. The pump polarization is rotated by a half-wave plate.
Pump and signal beams are temporally synchronized by a mechanically controlled  translation stage. 
The two beams are combined on a beam splitter (50:50 non-polarizing plate, Thorlabs BSW11R), then coupled to a 40X reflective objective with NA=0.5 via a second non-polarizing plate BSW11R. The spatial overlap at the sample plane is achieved by imaging the sample and the laser spots on a silicon camera.
The choice of a reflective objective is motivated by the achromatic focusing. The estimated spot sizes of pump and signal beams on the sample are $\sim$ 1$\mu$m and  $\sim$ 2
$\mu$m, respectively. 
In all the reported experiments the pump power ranges from 6 to 90$\mu$W, and the signal power ranges from 1 to 12.7mW. Both pump and signal powers are adjusted by variable neutral density attenuators. At the sample plane, if we account for the dispersion introduced by the crossed transmissive optics, pump and signal pulse duration is $\sim$250fs.
The detection process occurs in a back scattering geometry. The measured total collecton efficiency of the objective is 27.4\% and it is constant across the spectral bandwidth of our experiment (1.9-2.3eV). In the extracted absolute efficiency for the OPA process, we also consider that the collected idler power is actually half of the total emitted power. Due to the fact that the idler is generated in the nonlinear process, the reflected and transmitted idler powers coincide owing to the continuity of the electric field at the 1L-TMD plane, i.e., ${\bf E}_{\rm R}^{\rm (i)} = {\bf E}^{\rm (i)}_{\rm T}$ (no incident field oscillates at the idler frequency, ${\bf E}_0^{\rm (i)} = 0$). In the accounted losses of our system, we also include the 50:50 non-polarizing beam-splitter (Thorlabs BSW11R) transmission, for each idler photon energy and polarization, and the reflectivity of the two silver mirrors before the silicon CCD camera. The polarization- and grating-dependent count/photon ratio of the Silicon-EMCCD camera is carefully measured (see Supplementary Information) across a broad spectral range, fully covering our experimental working region, in order to directly convert the number of counts into number of photons, i.e., power.
After the interaction with the sample, in the collection path the input laser beams are filtered out by two 500nm (2.48eV) short-pass and 700nm (1.77eV) long-pass filters. In all the presented results pump and signal beams have linear polarizations, rotated by a zero-order 405nm (3.06eV) half-waveplate and a broadband (1100-2000nm / 0.62-1.13eV) half-waveplate, respectively. In order to directly compensate for all the polarization anisotropies of our setup, e.g., the different s and p polarization reflection and transmission from the beam splitter plates, the polarization-dependent measurements shown in Fig.\ref{Fig:emission} and Fig.\ref{Fig:emission}b are acquired in two configurations: 0$^\circ$ and 90$^\circ$. In the first configuration the sample is mounted at 0$^\circ$ with respect to the zig-zag direction, and the incoming polarization directions are set. In the second configuration the sample is mounted at 90$^\circ$ with respect to the zig-zag direction, and the incoming polarization directions are also rotated by 90$^\circ$ with respect to the ones set in the first configuration. The final measurement is the average of the two configurations. In Fig.\ref{Fig:emission}c the incoming polarization directions are set and fixed at 0$^\circ$. The sample is rotated by a rotational motor from 0$^\circ$ to 360$^\circ$ in steps of 5$^\circ$. The imaging systems enables to take every polar measurement on the same spot.\\

{\bf Parametric amplification by MX$_2$}.

Following a previously reported approach \cite{Ciattoni2018}, the linear and nonlinear surface conductivities of MX$_2$ are calculated from the tight-binding (TB) Hamiltonian of the electronic band structure \cite{LWY2013}, which is approximated as a ${\bf k}\cdot{\bf p}$ Hamiltonian $H_0({\bf k},\tau,s)$ (see Supplementary Information), where ${\bf k}$ is the electron wavenumber and $\tau$ and $s$ are the valley and spin indexes, respectively. Coupling with external radiation is introduced through the time-dependent Hamiltonian $H_0\left[{\bf k}+(e/\hbar){\bf A}(t),\tau,s\right]$, where $-e$ is the electron charge, $\hbar$ is the reduced Planck constant, and ${\bf A}(t)$ is the vector potential accounting for signal, idler, and pump fields with carrier angular frequencies $\omega_1$ ($\omega_{\rm s}$), $\omega_2$ ($\omega_{\rm i}$), and $\omega_3$ ($\omega_{\rm p}$), respectively. Such a time-dependent Hamiltonian leads to the temporal evolution of the density matrix $\dot{\rho} = -({\rm i}/\hbar)[H_0(t),\rho]-(1/\tau)(\rho-\rho_0)$, where a phenomenological relaxation is assumed to bring the system to the relaxed state $\rho_0$ at a rate $\tau^{-1}$ with $\hbar\tau^{-1}=50$\,meV (i.e., the relaxation time is taken as $\tau \approx 13$\,fs). We solve perturbatively the density-matrix equations of MX$_2$ in the weak excitation limit and in the slowly varying envelope approximation (see Supplementary Information), obtaining the surface current density 
\begin{eqnarray}
{\bf J}(t) & = & {\rm Re} \left\{ \sum_{j = 1}^3 \left [ \sigma_{\rm L} (\omega_j) A_j(t)\hat{\bf n}_j {\rm e}^{-i\omega_j t} \right] + \right. \nonumber \\ 
&& \left. + \sigma_2(\omega_1,\omega_2) A_1(t)A_2(t){\cal M} (\hat{\bf n}_1) \hat{\bf n}_2 {\rm e}^{-i\omega_3 t} + \right. \nonumber \\
&& \left.+ \sigma_2(\omega_1,\omega_3) A_1^*(t)A_3(t)\hat{\bf n} (\hat{\bf n}_1) \hat{\bf n}_3 {\rm e}^{-i\omega_2 t} + \right. \nonumber \\
&& \left. + \sigma_2(\omega_2,\omega_3)A_2^*(t)A_3(t) \hat{\bf n} (\hat{\bf n}_2) \hat{\bf n}_3 {\rm e}^{-i\omega_1 t} \right\},
\end{eqnarray}
\noindent where $\sigma_{\rm L}(\omega_j)$ ($j=1,2,3$) and $\sigma_2(\omega_l,\omega_m)$ ($l,m=1,2,3$) are the linear and nonlinear mixing conductivities, $\hat{\bf n}_j$ are in-plane linear polarization vectors $\hat{\bf n}_j = {\rm cos}\theta_j \hat{\bf x} + {\rm sin}\theta_j \hat{\bf y}$ defined in terms of angles $\theta_j$ with respect to the armchair edge (oriented along $\hat{\bf x}$),
\begin{eqnarray}
{\cal M}(\hat{\bf n}_j) = \left(\begin{array}{cc} \hat{\bf n}_j\cdot\hat{\bf y} & \hat{\bf n}_j\cdot\hat{\bf x} \\ \hat{\bf n}_j\cdot\hat{\bf x} & -\hat{\bf n}_j\cdot\hat{\bf y} \end{array}\right),
\end{eqnarray}
\noindent $\hat{\bf x}$, $\hat{\bf y}$ are orthogonal unit vectors spanning the MX$_2$ plane, and $A_j(t)$ are the in-plane signal ($j=1$), idler ($j=2$), and pump ($j=3$) field envelopes, respectively. We emphasize that the structure of the matrix $\hat{\bf n}(\hat{\bf n}_j)$ is a direct consequence of the $D_{\rm 3h}$ group symmetry of MX$_2$. 

By solving the nonlinear scattering of impinging signal ($j=1$) and pump ($j=3$) waves with amplitudes $A_1(t)=A_{\rm s}(t)$, $A_3(t)=A_{\rm p}(t)$, and polarization angles $\theta_1=\theta_{\rm s}$ and $\theta_3=\theta_{\rm p}$ in the undepleted pump approximation (see Supplementary Information), it is possible to calculate analytically the idler field polarization vector 
\begin{equation}
\hat{\bf n}_2 = {\rm cos}(\pi/2 - \theta_1 - \theta_3) \hat{\bf x} + {\rm sin}(\pi/2 - \theta_1 - \theta_3) \hat{\bf y},
\end{equation}
and the idler envelope 
\begin{equation}
A_2(t) = {\cal C}_1 A_1^*(t)A_3(t)/[{\cal C}_2-{\cal C}_3|A_3(t)|^2],
\end{equation} 
while we do not report here the cumbersome expressions for the coefficients ${\cal C}_1,{\cal C}_2,{\cal C}_3$ (see Supplementary Information). Note that in the weak excitation regime $|A_3(t)|^2<<|{\cal C}_2/{\cal C}_3|$ the predicted dependence of idler power on pump and signal power is linear $|A_2(t)|^2 \simeq |{\cal C}_1/{\cal C}_2|^2 |A_1(t)|^2|A_3(t)|^2$ and the theory well reproduces the experimental findings (see Fig. 1c). The emitted average idler powers reported in Fig. 1 are calculated as $P_{\rm i} = (1/2)\tau_{\rm rep}^{-1} \epsilon_0 c \int_{-\tau_{\rm rep}/2}^{\tau_{\rm rep}/2}|A_2(t)|^2dt$, where we have considered the laser repetition period $\tau_{\rm rep} = 12.5$ ns, input signal and pump envelopes $A_j(t) = {\cal A}_j {\rm exp}(-t^2/2\tau^2)$ (j=1,3) with duration $\tau = 250$ fs and amplitudes ${\cal A}_j = \sqrt{2I_j/\epsilon_0c}$, where 
$I_j = 4 {\rm log}2 P_j\tau/(\pi s_j^2\tau_{\rm rep})$ are peak intensities obtained by averaging over Gaussian spatial profiles with spots $s_1 = 2 \mu$m, $s_3 = 1 \mu$m, and $P_j$ are the impinging average powers of signal and idler waves. The polarization dependencies illustrated in Figs. 2,3 fully reproduce the analytically derived rule $\theta_{\rm i} = \pi/2 - \theta_{\rm s} - \theta_{\rm p}$. Finally, the effective bulk $\chi_2(\omega_l,\omega_m)$ responsible for the OPA process can be calculated as 
\begin{eqnarray}
\chi_2 (\omega_1,\omega_2) & = & i \sigma_2(\omega_1,\omega_2)/(t_{\rm 1L}\epsilon_0\omega_3), \nonumber \\
\chi_2 (\omega_2,\omega_3) & = & i \sigma_2(\omega_2,\omega_3)/(t_{\rm 1L}\epsilon_0\omega_1), \\
\chi_2 (\omega_1,\omega_3) & = & i \sigma_2(\omega_1,\omega_3)/(t_{\rm 1L}\epsilon_0\omega_2), \nonumber
\end{eqnarray}
where $t_{\rm 1L} = 0.65$ nm is the 1L thickness of MX$_2$ and $\epsilon_0$ is the dielectric permittivity of vacuum.

\end{document}